\documentstyle[aps,prc,twocolumn,epsf]{revtex}

\begin{document}
\title{RPA equations and the instantaneous Bethe-Salpeter equation}
\author{J.Resag, D.Sch\"utte}
\address{Institut f\"ur Theoretische Kernphysik,\\
         Universit\"at Bonn, Nussallee 14-16, 53115 Bonn, FRG}
\date{\today}
\maketitle

\begin{abstract}
We give a derivation of the particle-hole RPA equations for an
interacting multi-fermion system by applying the instantaneous
approximation to the amputated two-fermion propagator of the system.
In relativistic field theory the same approximation leads from the
fermion-antifermion Bethe-Salpeter equation to the Salpeter equation.
We show that RPA equations and Salpeter equation are indeed equivalent.
\end{abstract} \pacs{}

\narrowtext

\section{Introduction} \label{I}
In the study of systems of many interacting fermions one finds that
methods used in relativistic field theory usually correspond to
approximations well-known in nonrelativistic many-particle theory.
An interesting example is the correspondence of the Hartree-Fock
approximation and the so-called Gap-equation used e.g. in
the Nambu-Jona-Lasinio model \cite{NJL}, which both give a
first approximation for the many-body problem. An improvement is
obtained by including correlations in the excited states as well as
the ground state of the system. In nonrelativistic many-body theory
this leads to the particle-hole Random-Phase-Approximation (RPA)
equations.

The RPA equations have been derived by various methods
(see e.g. ref.\cite{RS}). The most systematic approach, however, is
the Green's function method, especially in view of possible
generalizations. Although the connection of the RPA equations to the
Bethe-Salpeter equation is mentioned in the literature \cite{FeWa}, a
clear and systematic description appears to be missing. It is the
purpose of this paper to close this gap and to establish the relation
to the Salpeter equation.

The paper is organized in the following way:
In Sec.\ref{II} we will give a derivation of the RPA equations
based on the Green's function method.
The derivation uses two approximations, i.e.
\begin{itemize}
\item
we apply the instantaneous approximation to the amputated two-fermion
propagator
\item
we substitute the full one-fermion propagators by the free ones.
\end{itemize}
In relativistic field theory the same approximations lead from
the fermion-antifermion Bethe-Salpeter equation \cite{BS,GL}
to the Salpeter equation \cite{Sa}. The structure of this
equation shows many similarities with the RPA equations (see e.g.
refs.\cite{La,RMMP}). We will show in Sec.\ref{III} that Salpeter
equation and RPA equations are indeed equivalent.
Concluding remarks are given in Sec.\ref{IV}.

\section{RPA equations and the instantaneous approximation} \label{II}
\subsection{Pole structure of the polarization propagator} \label{IIA}
Let \(H\) be the hamiltonian describing the dynamics of a
nonrelativistic or relativistic system
of many interacting identical fermions
with ground state \(|\psi_0\rangle\) and excited states
\(|\psi_{\lambda}\rangle,\;\lambda>0\). The corresponding energies
will be denoted as \(E_0\) and \(E_{\lambda}\)
with \(E_{\lambda} \le E_{\lambda'}\) for \(\lambda<\lambda'\).
For simplicity we assume a discrete energy spectrum, i.e. free states
are considered in a finite space volume.
In the relativistic case \(|\psi_0\rangle\) is usually the
vacuum so that the excited states
are particle-hole excitations of the vacuum,
e.g. mesons.

In order to keep the discussion complete and to introduce the notation
some well-known definitions and facts will be recalled in the following.

\begin{figure}
  \vspace{0.20in}
  \centering
  \leavevmode
  \epsfxsize=0.20\textwidth
  \epsffile{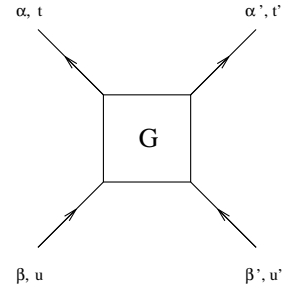}
  \vspace{0.20in}
\caption{The two-fermion propagator \(G\)}
\label{Gfig}
\end{figure}

\begin{figure}
  \vspace{0.20in}
  \centering
  \leavevmode
  \epsfxsize=0.40\textwidth
  \epsffile{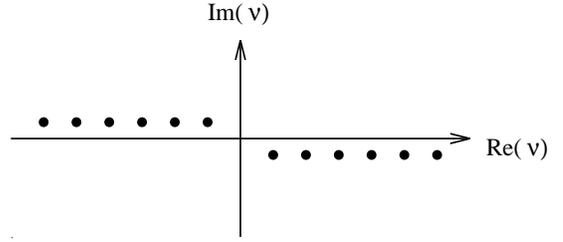}
  \vspace{0.20in}
\caption{Location of the poles of \(\Pi(\nu)\)}
\label{polesfig}
\end{figure}

Let \(a_{\alpha},\;a_{\alpha}^{\dagger}\) be fermion field operators
with the anticommutator given by
\( \left\{ a_{\alpha}^{\dagger},\,a_{\beta} \right\}
= \delta_{\alpha \beta} \). They correspond to an orthonormal
single particle basis
\(\varphi_{\alpha}\) which will be specified later.
The Heisenberg-picture will be used in the following and we define
\(A_{\alpha}(t) := e^{iHt}\,a_{\alpha}\,e^{-iHt}\). The two-fermion
propagator in this basis is then given by
\begin{eqnarray}
\lefteqn{i^2\,[G(t,t',u,u')]_{\alpha \alpha' \beta \beta'} =}
 \nonumber \\ &=&
\;\;\;\langle \psi_0 |
T\,A_{\alpha}(t)\,A_{\alpha'}(t')\,
A^{\dagger}_{\beta}(u)\,A^{\dagger}_{\beta'}(u')\,|
\psi_0 \rangle
 \nonumber \\ &=&
-\langle \psi_0 |
T\,A^{\dagger}_{\beta}(u)\,A_{\alpha}(t)\,
A^{\dagger}_{\beta'}(u')\,A_{\alpha'}(t')\,|
\psi_0 \rangle
\end{eqnarray}
(compare fig.\ref{Gfig}).
 \noindent
Let \(u=t+\epsilon\)
and \(u'=t'+\epsilon\) with \(\epsilon > 0\).
In the limit \(\epsilon \rightarrow 0\) one has
\begin{eqnarray}
\lefteqn{\lim_{\epsilon \rightarrow 0}\;
[G(t,t',t+\epsilon,t'+\epsilon)]
_{\alpha \alpha' \beta \beta'} =}
 \nonumber \\ &=&
\Theta(t-t')\,\langle \psi_0 |
\,A^{\dagger}_{\beta}(t)\,A_{\alpha}(t)\,
A^{\dagger}_{\beta'}(t')\,A_{\alpha'}(t')\,|
\psi_0 \rangle + \nonumber \\ &+&
\Theta(t'-t)\,\langle \psi_0 |
\,A^{\dagger}_{\beta'}(t')\,A_{\alpha'}(t')\,
A^{\dagger}_{\beta}(t)\,A_{\alpha}(t)\,|
\psi_0 \rangle
\end{eqnarray}
and with \(1 = \sum_{\lambda}\,|\psi_{\lambda}
\rangle \langle \psi_{\lambda}|\)
one obtains
\begin{eqnarray}
\lefteqn{\lim_{\epsilon \rightarrow 0}\;
[G(t,t',t+\epsilon,t'+\epsilon)]
_{\alpha \alpha' \beta \beta'} =}
 \nonumber \\ &=& \sum_{\lambda}\,\Big[
\Theta(t-t')\,e^{-i(E_{\lambda}-E_0)\,(t-t')}
 \nonumber \\ &&\;\;\;
\langle \psi_0         |a^{\dagger}_{\beta} 
a_{\alpha}| \psi_{\lambda} \rangle\,
\langle \psi_{\lambda} |a^{\dagger}_{\beta'}
a_{\alpha'}|\psi_0 \rangle \,
+ \nonumber \\
 && \;\;\;\;+ \;
\Theta(t'-t)\,e^{-i(E_{\lambda}-E_0)\,(t'-t)}
 \nonumber \\ &&\;\;\;
\langle \psi_0         |a^{\dagger}_{\beta'}
a_{\alpha'}|\psi_{\lambda} \rangle\,
\langle \psi_{\lambda} |a^{\dagger}_{\beta} 
a_{\alpha} |\psi_0 \rangle \,
\Big] \label{G}
\end{eqnarray}
The term for the ground state (\(\lambda=0\))
can be rewritten in terms of
one-fermion propagators
\(S^F_{\alpha\beta}(t-t')
= -i\,\langle \psi_0| T\,A_{\alpha}(t)\,
       A^{\dagger}_{\beta}(t')|\psi_0 \rangle \)
as 
\begin{eqnarray}
\lefteqn{
\langle \psi_0 |a^{\dagger}_{\beta} 
a_{\alpha} |\psi_0 \rangle\,
\langle \psi_0 |a^{\dagger}_{\beta'}
a_{\alpha'}|\psi_0 \rangle =} \nonumber \\
&=&
-\lim_{\epsilon \rightarrow 0}\,
S^F_{\alpha\beta}(-\epsilon)\, S^F_{\alpha'\beta'}(-\epsilon)
\end{eqnarray}
The polarization propagator \(\Pi(t-t')\) is now defined as
\begin{eqnarray}
\lefteqn{\lim_{\epsilon \rightarrow 0}\;
[G(t,t',t+\epsilon,t'+\epsilon)]
_{\alpha \alpha' \beta \beta'} =} \nonumber \\
&=&
-\lim_{\epsilon \rightarrow 0}\;
S^F_{\alpha\beta}(-\epsilon)\, S^F_{\alpha'\beta'}(-\epsilon)
+ i\, [\Pi(t-t')]_{\alpha \alpha' \beta \beta'}
\end{eqnarray}
so that
\begin{eqnarray}
\lefteqn{i\,[\Pi(t-t')]
_{\alpha \alpha' \beta \beta'} =}
 \nonumber \\ &=& \sum_{\lambda \not= 0}\,\Big[
\Theta(t-t')\,e^{-i(E_{\lambda}-E_0)\,(t-t')} \nonumber \\ &&\;\;\;
\langle \psi_0         |a^{\dagger}_{\beta} 
a_{\alpha}| \psi_{\lambda} \rangle\,
\langle \psi_{\lambda} |a^{\dagger}_{\beta'}
a_{\alpha'}|\psi_0 \rangle \,
+ \nonumber \\
 && \;\;\;\;+ \;
\Theta(t'-t)\,e^{-i(E_{\lambda}-E_0)\,(t'-t)} 
\nonumber \\ &&\;\;\;
\langle \psi_0         |a^{\dagger}_{\beta'}
a_{\alpha'}|\psi_{\lambda} \rangle\,
\langle \psi_{\lambda} |a^{\dagger}_{\beta} 
a_{\alpha} |\psi_0 \rangle \,
\Big]
\end{eqnarray}
With
\begin{equation}
\Theta(t)\,e^{-iEt} = \frac{1}{2\pi i}\,
\int \frac{e^{-i\nu t}\,d\nu}{E-\nu-i\epsilon}
\label{theta}
\end{equation}
one can write the fourier transform
\(\Pi(\nu)=\int dt\,e^{i\nu t}\,\Pi(t)\) as
\begin{eqnarray}
\lefteqn{[\Pi(\nu)]
_{\alpha \alpha' \beta \beta'} =}
 \nonumber \\ &=& \sum_{\lambda \not= 0}\,\Bigg[\,
\frac{
\langle \psi_0         |a^{\dagger}_{\beta} 
a_{\alpha}| \psi_{\lambda} \rangle\,
\langle \psi_{\lambda} |a^{\dagger}_{\beta'}
a_{\alpha'}|\psi_0 \rangle
}
{\nu-(E_{\lambda}-E_0-i\epsilon)}
 \nonumber \\ && \;\;\;\;- \;
\frac{
\langle \psi_0         |a^{\dagger}_{\beta'}
a_{\alpha'}|\psi_{\lambda} \rangle\,
\langle \psi_{\lambda} |a^{\dagger}_{\beta} 
a_{\alpha} |\psi_0 \rangle
}
{\nu+(E_{\lambda}-E_0-i\epsilon)}
\Bigg] \label{Pipoles}
\end{eqnarray}
Thus the energy levels of the excited states appear as a doubled
system of poles in the polarization propagator \(\Pi(\nu)\)
at \(\nu=\pm(E_{\lambda}-E_0-i\epsilon)\) as shown
in fig.\ref{polesfig}.
 \noindent
This doubling of the spectrum can be traced back to the
appearence of the time ordering operator \(T\) in the definition
of the two-fermion propagator \(G\). Any equation that looks for
the poles of \(\Pi\) will therefore obtain a doubled eigenvalue
spectrum. This statement holds for relativistic as well as
nonrelativistic calculations. It is clear that the appearence
of the second set of poles does not yield any further physical
information, since only half of the poles can be identified
with the eigenvalues of the hamiltonian \(H\).

\subsection{The instantaneous approximation} \label{IIB}
\begin{figure}
  \vspace{0.20in}
  \centering
  \leavevmode
  \epsfxsize=0.40\textwidth
  \epsffile{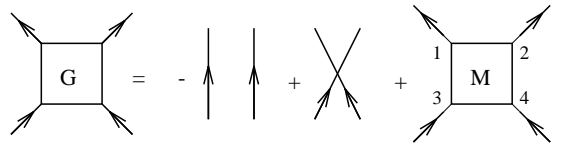}
  \vspace{0.20in}
\caption{Definition of the amputated two-fermion propagator \(M\)}
\label{Mfig}
\end{figure}

The amputated two-fermion propagator \(M\) is defined as
(see also fig.\ref{Mfig})
\begin{eqnarray}
\lefteqn{[G(t,t',u,u')]_{\alpha \alpha' \beta \beta'} =} \nonumber \\
& & - S^F_{\alpha\beta}(t-u)\, S^F_{\alpha'\beta'}(t'-u')
    + S^F_{\alpha\beta'}(t-u')\, S^F_{\alpha'\beta}(t'-u) + \nonumber
    \\
& & + \sum_{\alpha_1 \alpha_2} \sum_{\alpha_3 \alpha_4}\,
      \int dt_1 dt_2 dt_3 dt_4 \,
S^F_{\alpha\alpha_1}(t-t_1)\, S^F_{\alpha'\alpha_2}(t'-t_2) \cdot
\nonumber \\ && \;\;\cdot\;
[M(t_1,t_2,t_3,t_4)]_{\alpha_1 \alpha_2 \alpha_3 \alpha_4}\,
S^F_{\alpha_3\beta}(t_3-u)\, S^F_{\alpha_4\beta'}(t_4-u')
\end{eqnarray}
 \noindent
The instantaneous approximation of \(M\) used for
the investigation of one-particle -- one-hole propagation
is given by the ansatz
\begin{equation}
M \;\rightarrow\; M^{inst} = 
\delta(t_1-t_3)\,\delta(t_2-t_4)\,\Gamma(t_1-t_2)
\label{inst}
\end{equation}
i.e. it is assumed that particles and holes interact
instantaneously with each other.

\begin{figure}
  \vspace{0.20in}
  \centering
  \leavevmode
  \epsfxsize=0.40\textwidth
  \epsffile{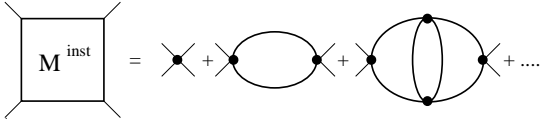}
  \vspace{0.20in}
\caption{Diagrams that contribute to \(M^{inst}\)
for a 4-point-interaction}
\label{MRPAfig}
\end{figure}

In nonrelativistic many-body theory this ansatz is equivalent to
taking into account only diagrams that have the appropriate
instantaneous structure, as shown in fig.\ref{MRPAfig}.
 \noindent
The same statement also holds for a relativistic 4-point-interaction.
In more general relativistic field theories like QCD,
however, the instantaneous approximation
cannot easily be interpreted in terms of Feynman diagrams.

With the relations of the previous section (using \(t'=0\))
the instantaneous approximation yields for the
polarization propagator
\begin{eqnarray}
\lefteqn{i\,[\Pi^{inst}(t)]_{\alpha \alpha' \beta \beta'} =
      S^F_{\alpha\beta'}(t)\, S^F_{\alpha'\beta}(-t) + }
\nonumber    \\
&&  + \sum_{\alpha_1 \alpha_2} 
\sum_{\alpha_3 \alpha_4}\,\int dt_1 dt_2\,
S^F_{\alpha\alpha_1}(t-t_1)\, S^F_{\alpha'\alpha_2}(-t_2)\cdot
\nonumber \\ && \;\;\;\;\;\;\;\;\;\;\cdot\;
[\Gamma(t_1-t_2)]_{\alpha_1 \alpha_2 \alpha_3 \alpha_4}\,
S^F_{\alpha_3\beta}(t_1-t)\, S^F_{\alpha_4\beta'}(t_2)
\label{PiRPA}
\end{eqnarray}
To simplify the notation we define
\begin{equation}
[g(t)]_{\alpha \alpha' \beta \beta'} :=
S^F_{\alpha\beta'}(t)\, S^F_{\alpha'\beta}(-t)
\end{equation}
Furthermore let
\begin{equation}
(A\,B)_{\alpha\beta\gamma\delta} := 
\sum_{\alpha' \beta'}\,A_{\alpha\alpha'\gamma\beta'}
                     \,B_{\beta'\beta\alpha'\delta}
\end{equation}
An easy way to represent this definition is to define
multi-indices as
\(
A_{\alpha\beta\gamma\delta} =:
A_{(\alpha\gamma)\,(\delta\beta)} =:
A_{ij} \label{multiindex}
\)
so that the usual matrix multiplication can be applied
to \(A_{ij}\).
With these definitions the Fourier transform
of eq.(\ref{PiRPA}) can be written as
\begin{equation}
i\,\Pi^{inst}(\nu) =
g(\nu) + g(\nu)\,\Gamma(\nu)\,g(\nu)
\end{equation}
The exact Bethe-Salpeter equation for the amputated
two-fermion propagator \(M\) reads
\begin{eqnarray}
M_{1234} &=& K_{1234} +
\sum_{\alpha_5 \alpha_6} \sum_{\alpha_7 \alpha_8}\,
\int dt_5\,dt_6\,dt_7\,dt_8
\nonumber \\
&& \left[
    K_{1537}\, S^F_{58}\, S^F_{67}\, M_{6284} +
    K_{2537}\, S^F_{58}\, S^F_{67}\, M_{6184} \right]
\end{eqnarray}
\begin{figure}
  \vspace{0.20in}
  \centering
  \leavevmode
  \epsfxsize=0.40\textwidth
  \epsffile{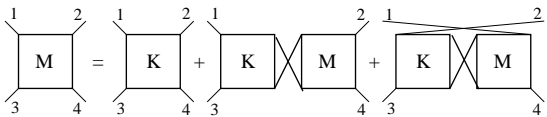}
  \vspace{0.20in}
\caption{The Bethe-Salpeter equation for the amputated two-fermion
propagator \(M\)}
\label{BSeqfig}
\end{figure} \noindent
(compare fig.\ref{BSeqfig}) with an appropriate particle-hole
kernel \(K\) and the notation
\([K(t_a,t_b,t_c,t_d)]_{\alpha_a\alpha_b\alpha_c\alpha_d}
=:K_{abcd}\) and analogously for \(M\) and \(S^F\).

In a first step one neglects the last (exchange) term.
The second step is to substitute \(M^{inst}\) for
\(M\) which implies that also \(K\)
must have this instantaneous structure, i.e.
\begin{equation}
K \;\rightarrow\; K^{inst} =
\delta(t_1-t_3)\,\delta(t_2-t_4)\,iV(t_1-t_2)
\end{equation}
After applying the Fourier transformation the instantaneous
Bethe-Salpeter equation reads
\begin{equation}
\Gamma(\nu) = iV(\nu) + iV(\nu)\,g(\nu)\,\Gamma(\nu)
\end{equation}
Together with \(i\,\Pi^{inst}=g+g\,\Gamma\,g\) one obtains
\(i\,\Pi^{inst}=g-g\,V\,\Pi^{inst}\) or
\begin{equation}
i\,\Pi^{inst}(\nu) = \left( [g(\nu)]^{-1}-i\,V(\nu) \right)^{-1}
\label{Piinv}
\end{equation}

\subsection{The RPA-equations}
From eq.(\ref{Pipoles}) we know that \(\Pi(\nu)\) has poles at
\(\nu=\pm(E_{\lambda}-E_0-i\epsilon)\). On the other hand
we can use eq.(\ref{Piinv}) to obtain a
spectral decomposition for \(\Pi^{inst}(\nu)\) in order to
calculate the pole positions and matrix elements in the
instantaneous approximation.
In the following we therefore look
for solutions \(F(\nu)\) of the equation
\begin{equation}
\left( [g(\nu)]^{-1}-i\,V(\nu) \right)\,F(\nu) = 0
\label{preRPA}
\end{equation}
To proceed further we approximate the full propagators \(S^F\)
in \(g\) by the free propagators
\(
[S^F_0(t)]_{\alpha\beta} = -i\,
\langle \chi_0 | T\,A^0_{\alpha}(t)\,
[A^0_{\beta}(0)]^{\dagger} |\chi_0
\rangle
\)
with
\(A^0_{\alpha}(t) = e^{iH_0t}\,a_{\alpha}\,e^{-iH_0t}\),
where
\(H_0 =
\sum_{\alpha}\,\epsilon_{\alpha}\,a^{\dagger}_{\alpha}a_{\alpha}\)
is some 'free' hamiltonian with ground state
\(|\chi_0\rangle\).
The basis states \(\varphi_{\alpha}\) have thus been chosen as
eigenstates of \(H_0\), i.e.
\(H_0\,\varphi_{\alpha} = \epsilon_{\alpha}\,\varphi_{\alpha}\).
In the nonrelativistic case an appropriate choice
for \(H_0\) is the Hartree-Fock hamiltonian, whereas in the
relativistic case one can use the free Dirac hamiltonian with some
effective fermion mass \(m\).

For a system of \(n\) fermions, the ground state \(|\chi_0\rangle\)
of \(H_0\) is a product wavefunction where the lowest \(n\)
eigenstates are occupied, i.e.
\(a_{\alpha}\,|\chi_0\rangle=0\) for \(\alpha>n\) and
\(a^{\dagger}_{\alpha}\,|\chi_0\rangle=0\) for \(\alpha \le n\).
In the relativistic case \(|\chi_0\rangle\) is the filled Dirac sea.
With
\begin{eqnarray}
H_0\,a^{\dagger}_{\alpha}\,|\chi_0\rangle &=&
( \epsilon_{\alpha}+\epsilon_0)\, 
a^{\dagger}_{\alpha}\,|\chi_0\rangle \\
H_0\,a_{\alpha}\,  |\chi_0\rangle &=&
(-\epsilon_{\alpha}+\epsilon_0)\, 
a_{\alpha}\,  |\chi_0\rangle
\end{eqnarray}
one finds
\begin{eqnarray}
\lefteqn{[iS^F_0(t)]_{\alpha\beta} =
e^{-i\epsilon_{\alpha}t}\,
\delta_{\alpha\beta}} \nonumber \\ && \cdot
\left[ \Theta(t)\,\Theta(\alpha-n) - 
\Theta(-t)\,\Theta(n-\alpha) \right]
\end{eqnarray}
and therefore
\begin{eqnarray}
\lefteqn{[g^0(t)]_{\alpha\alpha'\beta\beta'}
=[S^F_0(t)]_{\alpha\beta'}\,
[S^F_0(-t)]_{\alpha'\beta}= }\nonumber \\
&=& -e^{-(\epsilon_{\alpha}-\epsilon_{\beta})t}\,
\delta_{\alpha\beta'}\,\delta_{\alpha'\beta}\,\cdot
\nonumber \\ && \cdot
\big[ \Theta(t)\,\Theta(\alpha-n)\,\Theta(n-\beta)+
\nonumber \\ && +
      \Theta(-t)\,\Theta(\beta-n)\,\Theta(n-\alpha)
\big]
\end{eqnarray}
In the following we will use capital letters like \(\alpha=A\) for
\(\alpha>n\) and small letters like \(\alpha=a\) for \(\alpha \le n\).
Using eq.(\ref{theta}) for the \(\Theta\)-functions and the multi-index
notation of the previous section we find for the fourier transform
of \(g^0\)
\begin{eqnarray}
\left[\left(g^0(\nu)\right)^{-1}\right]_{(Ab)\,(B'a')}
&=& -i\,(\epsilon_A-\epsilon_b-\nu)\,\delta_{(Ab)\,(B'a')} \\
\left[\left(g^0(\nu)\right)^{-1}\right]_{(aB)\,(b'A')}
&=& -i\,(\epsilon_B-\epsilon_a+\nu)\,\delta_{(aB)\,(b'A')}
\end{eqnarray}
Note that the matrix elements of \(g\) are zero if one of
the multi-indices equals \( (ab)\) or \( (AB)\). Because of
\(i\Pi^{inst}=g+g\Gamma g\) the same holds for \(\Pi^{inst}\)
and because of \(i\Pi^{inst}=g-gV\Pi^{inst}\) also the matrix elements
of \(V\) with these indices are not relevant. It is therefore
sufficient to consider only the index combinations \( (Ab)\) and
\( (bA)\). This simplification is a consequence of
approximating the full fermion propagators by the free ones.
Write
\begin{equation}
F=
\left( \begin{array}{*{1}{c}}
X \\ Y
\end{array} \right)
\end{equation}
with \(X_k=F_{(Ab)}\) and \(Y_k=F_{(bA)}\). In the following we will
assume that \(X,\, Y\) are vectors of dimension \(D\), i.e.
\(k=1 \ldots D\). Define \(2D \times 2D\) -matrices
\(\Omega\) and \(N\) by
\([g^0(\nu)]^{-1} =: -i\,(\Omega-\nu\,N)\) so that
\begin{eqnarray}
\Omega_{(Ab)(B'a')} &=& (\epsilon_A-\epsilon_b)\,\delta_{(Ab)(B'a')}
\nonumber \\
\Omega_{(aB)(b'A')} &=& (\epsilon_B-\epsilon_a)\,\delta_{(aB)(b'A')}
\nonumber \\
     N_{(Ab)(B'a')} &=&  \delta_{(Ab)(B'a')}
\nonumber \\
     N_{(aB)(b'A')} &=& -\delta_{(aB)(b'A')}
\end{eqnarray}
which we will write in a simplified matrix notation as
\begin{equation}
\Omega=
\left( \begin{array}{*{2}{c}}
(\epsilon_A-\epsilon_b) & 0 \\
 0 & (\epsilon_B-\epsilon_a)
\end{array} \right)
\;\;\; , \;\;\;
N = 
\left( \begin{array}{*{2}{c}}
1 & 0 \\ 0 & -1
\end{array} \right)
\end{equation}
Setting \(h:=\Omega+V\) we find that
eq.(\ref{preRPA}) (multiplied with \(i\))
can be written as
\begin{equation}
h\,F^{\rho} = \nu_{\rho}\,N\,F^{\rho} \label{RPAeq}
\end{equation}
where \(\rho\) labels the different solutions and
\(F^{\rho}:=F(\nu_{\rho})\). 
In the nonrelativistic
case one usually approximates the kernel \(V\) by its lowest
order contribution, i.e. the two-fermion potential.
Then this equation is exactly the RPA-equation of e.g. ref.\cite{RS}.

Up to now it is not clear how to connect \(\nu_{\rho}\)
and \(F^{\rho}\) with the eigenvalues and eigenstates of the
full Hamiltonian \(H\). It is usefull at this point
to recall some of the properties of the RPA-equation:
\begin{itemize}
\item
\(h\) is a hermitian matrix and has the structure
\begin{equation}
h = 
\left( \begin{array}{*{2}{c}}
A & B \\ B^* & A^*
\end{array} \right)
\end{equation}
\item
Let
\begin{equation}
F^{\sigma} = 
\left( \begin{array}{*{1}{c}}
X^{\sigma} \\ Y^{\sigma}
\end{array} \right)
\end{equation}
be a solution with eigenvalue \(\nu_{\sigma}\). Then
\begin{equation}
F^{\tau} = 
\left( \begin{array}{*{1}{c}}
X^{\tau} \\ Y^{\tau}
\end{array} \right)
         = 
\left( \begin{array}{*{1}{c}}
(Y^{\sigma})^* \\ (X^{\sigma})^*
\end{array} \right)
\end{equation}
is a solution with eigenvalue \(\nu_{\tau}=-\nu_{\sigma}\)
(we assume that \(\nu_{\sigma}\) is real).
For the components of the eigenvectors this means that
\( F^{\tau}_{(Ab)}=[F^{\sigma}_{(bA)}]^*\) and
\( F^{\tau}_{(bA)}=[F^{\sigma}_{(Ab)}]^*\).
\item
If \(F^{\rho_1}\) and \(F^{\rho_2}\) are solutions with
\(\nu_{\rho_1} \not= \nu_{\rho_2}\) then
\begin{equation}
\langle F^{\rho_1} | N | \,F^{\rho_2} \rangle
 ~:= (F^{\rho_1})^{\dagger} \, N \,F^{\rho_2} = 0
\end{equation}
\item
If \(B\) is 'small enough' the \(2D\) eigenvalues \(\nu_{\rho}\)
are real and nonzero. We will use the index convention
\(\sigma=1 \;\ldots\; D\) for \(\nu_{\sigma}>0\)
and \(\tau=D+1 \;\ldots\;2D\) for \(\nu_{\tau}<0\)
with \(\nu_{\tau}=-\nu_{\sigma}\). The solutions can then be
normalized as
\begin{equation}
\langle F^{\rho'} | N | \,F^{\rho} \rangle =
N_{\rho}\,\delta_{\rho'\rho}
\end{equation}
with \(N_{\sigma}=1\) and \(N_{\tau}=-1\). They form a
\(2D\)-dimensional
basis, i.e. one can expand a vector \(F\) as
\(F=\sum_{\rho}\,c_{\rho}\,F^{\rho}\) with
\(c_{\rho}=N_{\rho}^{-1}\,\langle F^{\rho}|N|\,F \rangle\)
so that
\begin{equation}
1 = \sum_{\rho}\,N_{\rho}^{-1}\,|F^{\rho}\rangle \,\langle F^{\rho}|N| 
\end{equation}
\end{itemize}
The following calculation
\begin{eqnarray}
\lefteqn{(N\nu-h)^{-1}\,N|F^{\rho'}\rangle = } \nonumber \\
&=& (\nu-N^{-1}h)^{-1}\,|F^{\rho'}\rangle
 =  \frac{1}{\nu-\nu_{\rho'}}\,|F^{\rho'}\rangle = \nonumber \\
&=& \sum_{\rho}\,\frac{1}{\nu-\nu_{\rho}}
  \,N_{\rho}^{-1}\,|F^{\rho}\rangle \,\langle
  F^{\rho}|N|F^{\rho'}\rangle
\end{eqnarray}
now showes that the spectral decomposition of \((N\nu-h)^{-1}\)
is given by
\begin{equation}
(N\nu-h)^{-1} =
\sum_{\rho=1}^{2D}\,\frac{1}{\nu-\nu_{\rho}}
  \,N_{\rho}^{-1}\,|F^{\rho}\rangle \,\langle F^{\rho}|
\end{equation}
Let \( (\alpha\beta) \) stand for \((Ab)\) or \((aB)\).
Since \( F^{\tau}_{(\alpha\beta)}=[F^{\sigma}_{(\beta\alpha)}]^*\)
the spectral decomposition for the matrix elements
of \( \Pi^{inst}(\nu) \) can be written as
\begin{eqnarray}
\lefteqn{[\Pi^{inst}(\nu)]_{\alpha\delta\beta\gamma} =
(N\nu-h)^{-1}_{(\alpha\beta)\,(\gamma\delta)} = } \nonumber \\
&=& \sum_{\sigma=1}^{M}\,\left[
\frac{1}{\nu-\nu_{\sigma}}
\,F^{\sigma}_{(\alpha\beta)}\,[F^{\sigma}_{(\gamma\delta)}]^*
~- \frac{1}{\nu+\nu_{\sigma}}
\,[F^{\sigma}_{(\beta\alpha)}]^*\,F^{\sigma}_{(\delta\gamma)}
\,\right]
\end{eqnarray}
Comparing this equation with the exact pole structure given in
eq.(\ref{Pipoles}) we can identify
\begin{eqnarray}
F^{\sigma}_{\alpha\beta} &=&
\langle \psi_0         |a^{\dagger}_{\beta} 
a_{\alpha}| \psi_{\sigma} \rangle
\\
\nu_{\sigma} &=& E_{\sigma}-E_0
\end{eqnarray}

\section{From the Salpeter equation to the RPA-equations} \label{III}
In relativistic field theory particle-hole excitations of the
fermionic vacuum are described by the fermion-antifermion Bethe-Salpeter
equation \cite{BS,GL}. In the instantaneous approximation this
equation reduces to the Salpeter equation \cite{Sa}. From the
considerations of the previous section we expect the Salpeter equation
to be equivalent to the RPA equations (\ref{RPAeq}). We will show in this
section that this is indeed the case.

The Salpeter equation for one fermion flavor
can be written in the form (see e.g. \cite{La,RMMP})
\begin{equation}
({\cal H}\psi)(\vec{p}) = M\,\psi(\vec{p})
\label{19} \end{equation}
where
\begin{eqnarray}
\lefteqn{ ({\cal H}\psi)(\vec{p})
 = H(\vec{p}) \psi(\vec{p}) - \psi(\vec{p}) H(\vec{p})} \nonumber \\ &&
 -\int \frac{d^3p'}{(2\pi)^3}  \,
 \Lambda^+(\vec{p}) \,[W(\vec{p},\vec{p}\,')\, \psi(\vec{p}\,')]\,
\Lambda^-(\vec{p}) \nonumber \\ &&
 +\int \frac{d^3p'}{(2\pi)^3}  \,
 \Lambda^-(\vec{p}) \,[W(\vec{p},\vec{p}\,')\, \psi(\vec{p}\,')]\,
\Lambda^+(\vec{p}) \label{20}
\end{eqnarray}
with the free Dirac hamiltonian 
\(H(\vec{p})=\gamma^0\,(\vec{\gamma}\vec{p}+m)\),
the projection operators \(\Lambda^{\pm}=(\omega 
\pm H(\vec{p}))/(2\omega)\) and
\(\omega=\sqrt{\vec{p}^{\,2}+m^2}\) 
(don't confuse \(H(\vec{p})\) with
the full hamiltonian of the previous section). 
Here \(M\) is the mass of the bound state, \(m\) is the effective
fermion mass and \(W\) is the instantaneous interaction kernel.
One can define a scalar product by
\begin{equation}
\left\langle \psi_1 \right.\left| \psi_2 \right\rangle
= \int\,\mbox{tr}\,\left( 
    \psi_1^{\dagger}\,\Lambda^+\,
    \psi_2\,\Lambda^- -
    \psi_1^{\dagger}\,\Lambda^-\,
    \psi_2\,\Lambda^+ \right) \nonumber \\
\end{equation}
with all quantities depending on \(\vec{p}\) and the notation
\(\int=\int\,d^3p/(2\pi)^3\).

Let \(u(\vec{p}),\,v(\vec{p})\) be free
Dirac spinors (we use the conventions of \cite{IZ} in the following).
Define
\begin{eqnarray}
w^{(+)}_{rs}(\vec{p})
&:=& u_r(\vec{p}) \otimes v_s^{\dagger}(-\vec{p})
\nonumber \\
w^{(-)}_{rs}(\vec{p})
&:=& v_s(-\vec{p}) \otimes u_r^{\dagger}(\vec{p})
\end{eqnarray}
We will use the box normalization in the following, i.e. we substitute
\begin{equation}
\int \frac{d^3p}{(2\pi)^3} \;\longrightarrow\;
\frac{1}{V}\,\sum_{\vec{p}}
\end{equation}
Since
\(\Lambda^+ \psi \,\Lambda^+ = \Lambda^- \psi \,\Lambda^- = 0\)
we can expand
\begin{equation}
\psi(\vec{p}) = \sqrt{V}\,\frac{m}{\omega}\,
\sum_{r,s=\pm 1/2}\,
\left( b^{(+)}_{\vec{p},rs}\,w^{(+)}_{rs}(\vec{p}) +
       b^{(-)}_{\vec{p},rs}\,w^{(-)}_{rs}(\vec{p}) \right)
\label{exppsip}
\end{equation}
with some suitable coefficients \(b\). The factors in front
of the summation sign have been choosen to simplify the notation
in the following.

Solving the Salpeter equation \({\cal H}\psi=M\,\psi\)
is equivalent to solving
\begin{equation}
\langle \psi_1|{\cal H}\,\psi_2 \rangle
= M\,\langle \psi_1|\psi_2 \rangle
\end{equation}
for all given \(\psi_1\).

With eq.(\ref{exppsip}) and the relations
for \(u,\,v\) of ref.\cite{IZ} we compute
\begin{eqnarray}
&&\mbox{tr}\,
\left[
[w^{(+)}_{rs}]^{\dagger}\,\Lambda^+\,
w^{(+)}_{r's'}\,\Lambda^- -
[w^{(+)}_{rs}]^{\dagger}\,\Lambda^-\,
w^{(+)}_{r's'}\,\Lambda^+
\right] = \nonumber \\
&=& +\frac{\omega^2}{m^2}\,\delta_{rr'}\,\delta_{ss'} \\
&&\mbox{tr}\,
\left[
[w^{(-)}_{rs}]^{\dagger}\,\Lambda^+\,
w^{(-)}_{r's'}\,\Lambda^- -
[w^{(-)}_{rs}]^{\dagger}\,\Lambda^-\,
w^{(-)}_{r's'}\,\Lambda^+
\right] = \nonumber \\
&=& -\frac{\omega^2}{m^2}\,\delta_{rr'}\,\delta_{ss'} \\
&&\mbox{tr}\,
\left[
[w^{(+)}_{rs}]^{\dagger}\,\Lambda^+\,
w^{(-)}_{r's'}\,\Lambda^- -
[w^{(+)}_{rs}]^{\dagger}\,\Lambda^-\,
w^{(-)}_{r's'}\,\Lambda^+
\right] = \nonumber \\
&=& 0 \\
&&\mbox{tr}\,
\left[
[w^{(-)}_{rs}]^{\dagger}\,\Lambda^+\,
w^{(+)}_{r's'}\,\Lambda^- -
[w^{(-)}_{rs}]^{\dagger}\,\Lambda^-\,
w^{(+)}_{r's'}\,\Lambda^+
\right] = \nonumber \\
&=& 0
\end{eqnarray}
where all quantities depend on \(\vec{p}\).
Using the multiindex \(i=(\vec{p},r,s)\)
the scalar product in the box normalization can therefore be written as
\begin{eqnarray}
\langle \psi_1 | \psi_2 \rangle &=&
\sum_i\,\left[
(b_1^{(+)})_i^* \,(b_2^{(+)})_i - (b_1^{(-)})_i^* \,(b_2^{(-)})_i
\right] = \nonumber \\ &=&
\left( \begin{array}{*{1}{c}}
b_1^{(+)} \\ b_1^{(-)}
\end{array} \right)^{\dagger}\,
\left( \begin{array}{*{2}{c}}
1 & 0 \\ 0 &-1
\end{array} \right)\,
\left( \begin{array}{*{1}{c}}
b_2^{(+)} \\ b_2^{(-)}
\end{array} \right)
\end{eqnarray}
We further have
\begin{eqnarray}
\langle \psi_1| {\cal H}\,\psi_2 \rangle &=&
\langle \psi_1| {\cal T}\,\psi_2 \rangle  +
\langle \psi_1| {\cal V}\,\psi_2 \rangle \;\;\;\mbox{with}\\
\langle \psi_1| {\cal T}\,\psi_2 \rangle &=&
\frac{1}{V}\,\sum_{\vec{p}}\,2\omega\,
\mbox{tr}\,\left(\psi_1^{\dagger}(\vec{p})\,\psi_2(\vec{p})\right) \\
\langle \psi_1| {\cal V}\,\psi_2 \rangle &=&
-\frac{1}{V^2}\,\sum_{\vec{p}}\,\sum_{\vec{p}\,'}\,
\mbox{tr}\,\left(\psi_1^{\dagger}(\vec{p})\,
W(\vec{p},\vec{p}\,')\,\psi_2(\vec{p}\,')\right)
\end{eqnarray}
We proceed analogously for the kinetic energy term
and compute
\begin{equation}
\mbox{tr}\,\left[
[w^{(a)}_{rs}(\vec{p})]^{\dagger}\,[w^{(a')}_{r's'}(\vec{p})]
\right]
= \frac{\omega^2}{m^2}\,\delta_{rr'}\,\delta_{ss'}
  \,\delta_{aa'}
\end{equation}
(where \(a,a'=\pm\)) so that with \(\omega_i=\omega(\vec{p})\)
\begin{eqnarray}
\langle \psi_1| {\cal T}\,\psi_2 \rangle  &=&
\sum_i\,2\omega_i\,\left[
(b_1^{(+)})_i^* \,(b_2^{(+)})_i + (b_1^{(-)})_i^* \,(b_2^{(-)})_i
\right] = \nonumber \\ &=:&
\left( \begin{array}{*{1}{c}}
b_1^{(+)} \\ b_1^{(-)}
\end{array} \right)^{\dagger}\,
\left( \begin{array}{*{2}{c}}
2\omega & 0 \\ 0 & 2\omega
\end{array} \right)\,
\left( \begin{array}{*{1}{c}}
b_2^{(+)} \\ b_2^{(-)}
\end{array} \right)
\end{eqnarray}
For the interaction term we define
\begin{equation}
V^{a_1a_2}_{ij} :=
-\frac{1}{V}
\,\frac{m}{\omega}
\,\frac{m}{\omega'}
\,\mbox{tr}\,\left( [w_{r_1s_1}^{(a_1)}(\vec{p})]^{\dagger}
\,W(\vec{p},\vec{p}\,')\,w_{r_2s_2}^{(a_2)}(\vec{p}\,') \right)
\end{equation}
with \(a_1,a_2=\pm\) and the multiindices
\(i=(\vec{p},r_1,s_1),\;j=(\vec{p}\,',r_2,s_2)\).
Consider interaction kernels that fulfill the relation
\( [W(\vec{p},\vec{p}\,')\,\psi(\vec{p}\,')]^{\dagger} =
    W(\vec{p},\vec{p}\,')\,[\psi(\vec{p}\,')]^{\dagger} \)
(this is usually the case for kernels of physical interest).
Since \([w_{rs}^{(+)}(\vec{p})]^{\dagger} = w_{rs}^{(-)}(\vec{p})\)
we have \( V_{ij}^{--}=(V_{ij}^{++})^* \) and
        \( V_{ij}^{-+}=(V_{ij}^{+-})^* \)
so that we can write
\begin{eqnarray}
\langle \psi_1| {\cal V}\,\psi_2 \rangle  &=&
\sum_{i,j}\,\sum_{a_1,a_2=\pm}\,
(b_1^{(a_1)})_i^* \, V^{a_1a_2}_{ij}\,(b_2^{(a_2)})_j
\nonumber \\ &=&
\left( \begin{array}{*{1}{c}}
b_1^{(+)} \\ b_1^{(-)}
\end{array} \right)^{\dagger}\,
\left( \begin{array}{*{2}{c}}
V^{++} & V^{+-} \\ (V^{+-})^* & (V^{++})^*
\end{array} \right)\,
\left( \begin{array}{*{1}{c}}
b_2^{(+)} \\ b_2^{(-)}
\end{array} \right)
\end{eqnarray}
Since \(b_1^{(\pm)}\) are arbitrary
the Salpeter equation can now be written in matrix form as
\begin{eqnarray}
&&\left[
\left( \begin{array}{*{2}{c}}
2\omega &  0 \\ 0 & 2\omega
\end{array} \right)
+
\left( \begin{array}{*{2}{c}}
V^{++} & V^{+-} \\ (V^{+-})^* & (V^{++})^*
\end{array} \right)\,\right]
\left( \begin{array}{*{1}{c}}
b^{(+)} \\ b^{(-)}
\end{array} \right) =
\nonumber \\
&=& M\;
\left( \begin{array}{*{2}{c}}
 1 & 0 \\ 0 & -1
\end{array} \right)\,
\left( \begin{array}{*{1}{c}}
b^{(+)} \\ b^{(-)}
\end{array} \right)
\label{Salp}
\end{eqnarray}
We identify the eigenvalues of the free hamiltonian
with the kinetic energies of the free fermions as
\(\epsilon_A = -\epsilon_a = \omega_i \) and
\(\epsilon_B = -\epsilon_b = \omega_i \).
Further we identify the positive eigenvalues \(M\)
with \(\nu_{\sigma}=E_{\sigma}-E_0\)
and
\begin{equation}
\left( \begin{array}{*{1}{c}}
b^{(+)} \\ b^{(-)}
\end{array} \right)
 = F^{\sigma} =
\left( \begin{array}{*{1}{c}}
X^{\sigma} \\ Y^{\sigma}
\end{array} \right)
\end{equation}
Therefore we find that the Salpeter equation (\ref{Salp}) has exactly the
form of the RPA-equations (\ref{RPAeq}).

In this context we would like to mention a work of J.Piekarewicz
\cite{Pie} which gives a direct derivation of the Salpeter equation
using a method similar to our derivation of the RPA equations as given
in section \ref{II}.

\section{Conclusion} \label{IV}
In the present paper we have shown that the RPA equations can be
derived by applying the instantaneous approximation to the amputated
two-fermion propagator and by approximating the full fermion
propagators by the free ones. Our derivation holds for nonrelativistic
as well as relativistic fermionic systems. Since in relativistic field
theory the same approximations lead from the fermion-antifermion
Bethe-Salpeter equation to the Salpeter equation, this equation should
be equivalent to the RPA equations. We have shown explicitely that
this is indeed the case.

The RPA equations have been analyzed carefully by many authors,
especially in the context of nuclear physics (compare e.g. the
references given in \cite{RS}). It would be interesting to transfer
these results to the Salpeter equation, e.g. results
on the stability of the RPA or on the appearence of spurious
solutions.


\begin{thebibliography}{99}
\bibitem{NJL}  Y.Nambu, G.Jona-Lasinio, Phys.Rev. 122, (1961) 345
\bibitem{RS}   P.Ring, P.Schuck, The Nuclear Many-Body Problem
                   (Springer-Verlag, Berlin, 1980)
\bibitem{FeWa} A.L.Fetter, J.D.Walecka,
               Quantum Theory of Many Particle Systems
                   (McGraw-Hill, New York, 1971)
\bibitem{BS}   E.E.Salpeter, H.A.Bethe, Phys.Rev. 84, (1951) 132
\bibitem{GL}   M.Gell-Mann, F.Low, Phys.Rev. 84, (1951) 350
\bibitem{Sa}   E.E.Salpeter, Phys.Rev. 87, (1952) 328
\bibitem{La}   J.F.Laga\"e, Phys.Rev. D45, (1992) 305, 317
\bibitem{RMMP} J.Resag, C.R.M\"unz, B.C.Metsch, H.R.Petry,
                   Institut f\"ur theor. Kernphysik der Univ. Bonn,
                   preprint TK-93-13 (1993), nucl-th/9307026
\bibitem{IZ}   C.Itzykson, J.-B.Zuber, Quantum Field Theory
                   (McGraw-Hill, New York, 1985)
\bibitem{Pie}  J.Piekarewicz, Revista Mexicana de F\'isica 39,
                   No.4, (1993) 542
\end{thebibliography}
\end{document}